# Annular Bragg Defect mode Resonators

Jacob Scheuer and Amnon Yariv

Department of applied Physics, 128-95 California Institute of Technology,

Pasadena, California 91125

koby@caltech.edu

**Abstract**

We propose and analyze a new type of a resonator in an annular geometry which is based on a single defect surrounded by radial Bragg reflectors on both sides. We show that the conditions for efficient mode confinement are different from those of the conventional Bragg waveguiding in a rectangular geometry. A simple and intuitive approach to the design of optimal radial Bragg reflectors is proposed and employed, yielding chirped gratings. Small bending radii and strong control over the resonator dispersion are possible by the Bragg confinement. A design compromise between large Free Spectral Range (FSR) requirements and fabrication tolerances is suggested.

## 1. Introduction

The last few years, have witnessed a substantial increase of activity dealing with utilization of ring resonators for optical communication devices. Various ring-resonator based applications such as modulators [1], channel drop filters [2] and dispersion compensators [3] have been suggested and demonstrated.

The important characteristics of the modes of ring resonators are the free spectral range (FSR) and the loss per revolution, or equivalently the Q-factor. One method for realizing tight confinement and high Q is to utilize Bragg reflection instead of total internal reflection (as in "conventional" resonators"). Disk resonators based on Bragg reflection were analyzed in the past, both for laser and passive resonator applications [4-12], employing both coupled mode theory and field transfer matrices.

In this paper we propose and analyze a new type of ring resonator - an annular defect mode resonator which is based on a single annular defect located between radial Bragg reflectors. Bragg reflection based disk (i.e., a disk surrounded by concentric Bragg layers) and, recently, ring resonators have been studied theoretically and demonstrated experimentally [4-13]. Recently, a hexagonal waveguide ring resonator based on Photonic Bandgap Crystal (PBC) confinement on both sides of the waveguide was demonstrated experimentally [14]. However, this structure exploited the specific symmetry of the triangular lattice which enables low loss 60° abrupt turns in order to realized a closed resonator.

The basic geometry is illustrated in Fig. 1. A circumferentially guiding defect is located within a medium which consists of annular Bragg layers. Due to the circular geometry the layer thicknesses, unlike in rectangular geometry, are not constant [15] and our task is to determine the thicknesses which lead to maximum confinement in the defect.

In section 2 we develop the matrix formalism for solving for the modal field and in section 3 we describe the rules for designing an annular Bragg defect mode resonator. In section 4 we describe the dispersion relation and the modal profile of the field. In section 5 we analyze the properties of a resonators which are based on higher Bragg order reflectors and in section 6 we discuss the results and summarize.

## 2. Basic theory

We consider an azimuthally symmetric structure as illustrated in figure 1. The guiding defect which is comprised of a material with refractive index $n_{defect}$ is surrounded by distributed Bragg reflectors on both sides where the reflectors' layers are of refractive indices $n_1$ and $n_2$. All the electromagnetic field components can be expressed by the $z$ component of the electrical and magnetic field [15] which satisfy the Helmholtz equations which in cylindrical coordinates are given by:

$$\left[\frac{1}{\rho}\frac{\partial}{\partial\rho}\left(\rho\frac{\partial}{\partial\rho}\right)+\frac{1}{\rho^2}\frac{\partial^2}{\partial\theta^2}+k_0^2n^2(\rho)+\frac{\partial^2}{\partial z^2}\right]\begin{pmatrix}E_z\\ H_z\end{pmatrix}=0 \qquad (1)$$

where $\rho$, $z$ and $\theta$ are the radial, axial and azimuthal coordinates respectively and $k_0$ is the wavenumber in vacuum. The refractive index $n(\rho)$ equals either $n_{defect}$, $n_1$ or $n_2$ according to the radius $\rho$. Assuming that the $\rho$, $\theta$ and $z$ dependence of the field can be separated, the electrical $z$ field component can be written as:

$$E_z = R(\rho)\cdot\exp[i(m\theta\pm\beta z)] \quad m \text{ is an integer} \qquad (2)$$

with a similar expression for the magnetic field $z$ component. Introducing (2) into (1) leads to:

$$\rho^2 \frac{\partial^2 R}{\partial \rho^2} + \rho \frac{\partial R}{\partial \rho} + \left[\left(k^2(\rho) - \beta^2\right)\rho^2 - m^2\right]R = 0 \tag{3}$$

where $k(\rho) = k_0 n(\rho)$ is constant in each layer. The general solution of equation (3) can be expressed by a superposition of the Bessel functions of the first and second kind:

$$R_m(\rho) = A \cdot J_m\left(\sqrt{k_j^2 - \beta^2}\,\rho\right) + B \cdot Y_m\left(\sqrt{k_j^2 - \beta^2}\,\rho\right) \tag{4}$$

where $k_j$ is the material wavenumber in the $j^{\text{th}}$ layer. Combining equations (3-4), the electrical and magnetic $z$ components of the field are given by:

$$\begin{aligned} E_z &= \left[A \cdot J_m\left(\sqrt{k_j^2 - \beta^2}\,\rho\right) + B \cdot Y_m\left(\sqrt{k_j^2 - \beta^2}\,\rho\right)\right] \cdot \cos(\beta \cdot z + \varphi) \cdot \exp(im\theta) \\ H_z &= \left[C \cdot J_m\left(\sqrt{k_j^2 - \beta^2}\,\rho\right) + D \cdot Y_m\left(\sqrt{k_j^2 - \beta^2}\,\rho\right)\right] \cdot \sin(\beta \cdot z + \varphi) \cdot \exp(im\theta) \end{aligned} \tag{5}$$

The other fields' components are derived from $E_z$ and $H_z$:

$$\begin{aligned} E_\theta &= \frac{i}{\gamma_j^2} \cdot \left(\omega\mu \frac{\partial H_z}{\partial \rho} - \frac{m}{\rho}\frac{\partial E_z}{\partial z}\right) \\ E_\rho &= \frac{1}{\gamma_j^2} \cdot \left(\frac{\partial^2 E_z}{\partial z \partial \rho} - \frac{m\omega\mu}{\rho} H_z\right) \\ H_\theta &= \frac{-i}{\gamma_j^2} \cdot \left(\omega\varepsilon \frac{\partial E_z}{\partial \rho} + \frac{m}{\rho}\frac{\partial H_z}{\partial z}\right) \\ H_\rho &= \frac{1}{\gamma_j^2} \cdot \left(\frac{\partial^2 H_z}{\partial z \partial \rho} + \frac{m\omega\varepsilon}{\rho} E_z\right) \end{aligned} \tag{6}$$

Where $\gamma_j = \sqrt{k_j^2 - \beta^2}$ and $\mu$ and $\varepsilon$ are the dielectric and magnetic susceptibilities respectively.

Introducing (5) into (6) yields all the fields' components in the $j^{\text{th}}$ layer. The parallel component of the fields - $E_z, H_z, E_\theta, H_\theta$ must be continuous at the interfaces. This requirement can be written in from of a transfer matrix, connecting the amplitude vector $\begin{bmatrix} A & B & C & D \end{bmatrix}$ in the $j^{\text{th}}$ and $j$+1 layers:

$$\begin{pmatrix} E_z \\ H_\theta \\ H_z \\ E_\theta \end{pmatrix} = \begin{pmatrix} J(\gamma_j \rho) & Y(\gamma_j \rho) & 0 & 0 \\ \frac{n_j^2}{\gamma_j} J'(\gamma_j \rho) & \frac{n_j^2}{\gamma_j} Y'(\gamma_j \rho) & \frac{m\beta}{\rho\omega\varepsilon_0\gamma_j^2} J(\gamma_j \rho) & \frac{m\beta}{\rho\omega\varepsilon_0\gamma_j^2} Y(\gamma_j \rho) \\ 0 & 0 & J(\gamma_j \rho) & Y(\gamma_j \rho) \\ \frac{m\beta}{\rho\omega\mu\gamma_j^2} J(\gamma_j \rho) & \frac{m\beta}{\rho\omega\mu\gamma_j^2} Y(\gamma_j \rho) & \frac{1}{\gamma_j} J'(\gamma_j \rho) & \frac{1}{\gamma_j} Y'(\gamma_j \rho) \end{pmatrix} \begin{pmatrix} A_j \\ B_j \\ C_j \\ D_j \end{pmatrix} \equiv \tag{7}$$

$$\equiv \overline{\overline{M}}_j(\rho) \begin{pmatrix} A_j \\ B_j \\ C_j \\ D_j \end{pmatrix}$$

where $\omega$ is the optical angular frequency and the prime indicates a derivative with respect to the function argument and $\overline{\overline{M}}_j$ is the matrix to the right of the equality sign. The continuity consideration of the tangential electric and magnetic fields at the boundary $\rho_j$ separating the boundary layer $j$ and $j$+1 leads to:

$$\begin{pmatrix} A_{j+1} \\ B_{j+1} \\ C_{j+1} \\ D_{j+1} \end{pmatrix} = \overline{\overline{M}}_{j+1}^{-1}(\rho_j) \cdot \overline{\overline{M}}_j(\rho_j) \cdot \begin{pmatrix} A_j \\ B_j \\ C_j \\ D_j \end{pmatrix} \tag{8}$$

It is obvious from the structure of $M_j$ that the natural polarizations of the structure are not pure TE or TM. In this paper we are interested primarily in ring resonators modes such that $\beta \approx 0$. In this case equations (6) and (7) admit two independent types of solutions – a TE mode with $E_z$, $H_\rho$ and $H_\theta$ and a TM mode with $H_z$, $E_\rho$ and $E_\theta$.

We consider the "TE" component of the electromagnetic field which is characterized by $E_z$, $H_\rho$ and $H_\theta$. We designate this component as "TE" because the primary direction of the propagation is $\theta$. The $M$ matrix for this component is given by (7) with $\beta \approx 0$:

$$M_j = \begin{pmatrix} J(\gamma_j \rho) & Y(\gamma_j \rho) \\ \frac{n_j^2}{\gamma_j} J'(\gamma_j \rho) & \frac{n_j^2}{\gamma_j} Y'(\gamma_j \rho) \end{pmatrix} \tag{9}$$

Using relation (8), the field components $A$ and $B$ can be "propagated" from the inner layers to the external layers. We use the finiteness of the field at $\rho$=0 so that $B_1$=0. The second boundary condition is that past the last layer there is no inward propagating field so that $B_{N+1} = -iA_{N+1}$ (for the TE mode) and $N$ is the number of layers.

The employment of the transfer matrices is important here because, in contrast to coupled mode theory [5, 7], it enables an exact analysis of high-contrast Bragg structures that cannot be considered as small perturbation.

## 3. Design rules

The formalism of section 2 enables us to find the modal field distribution in the case of an arbitrary arrangement of annular concentric dielectric rings. We are especially interested in structures that can lead to a concentration of the modal energy near a predetermined radial distance i.e., in a radial defect mode.

High efficiency Bragg reflectors in Cartesian coordinates require a constant grating period which determines the angles in which an incident wave would be reflected. Generally, the grating wavenumber ($2\pi/\Lambda$ where $\Lambda$ is the grating period) multiplied by the reflection order should be approximately twice the transverse component of the incident wave's wavevector [15]. However, when the structure is annular, the conditions for efficient reflection are different.

Several methods for determining the thickness, and thus the position, of the Bragg layers interfaces have been suggested in previous publications [5-8]. Compared to Bragg fibers [16], the incident angle of the waves at the interfaces (measured from the normal to the interface) is smaller and, therefore, the asymptotic approximation [17] is not valid and the "conventional" $\lambda/4$ layers would not be appropriate. The principle underlying these methods is to position the layers interfaces at the zeros and extrema of the field transverse profile. This strategy ensures the decrease of the field intensity for larger radii and the reduction of radiating power from the resonator. Here we present a more intuitive, although equivalent, approach to determine the width of the layers.

We use the following conformal transformation [18, 19]:

$$\begin{aligned} \rho &= R \cdot \exp(U/R) \\ \theta &= V/R \end{aligned} \tag{10}$$

And the inverse transformation:

$$
\begin{aligned}
&U = R \cdot \ln(\rho/R) \\
&V = \theta \cdot R \\
&n(\rho) = n_{eq}(\rho) \cdot R/\rho
\end{aligned}
\tag{11}
$$

where $R$ is an arbitrary parameter. The transformation (10) maps a circle in the ($\rho$, $\theta$) plane with radius $R_0$ to a straight line in the ($U$, $V$) plane located at $U_0 = R \cdot \ln(R_0/R)$. The structure in figure 1 is transformed into a series of straight lines. The wave equation in the ($U$, $V$) plane is obtained by transforming (1):

$$
\frac{\partial^2 E}{\partial U^2} + \frac{\partial^2 E}{\partial V^2} + k_0^2 n_{eq}^2(U) E = 0 \tag{12}
$$

where $n_{eq}(U) = n(U) \cdot \exp(U/R)$ is the profile of the refractive index in the ($U$, $V$) plane. Fig. 2B depicts the equivalent index profile, $n_{eq}(U)$, in the ($U$, $V$) plane corresponding to the real index profile $n(\rho)$ shown in Fig. 2A. The later exemplifies a "conventional" Bragg waveguide design comprising $\lambda/4$ layers of alternating refractive indices and a $\lambda/2$ defect. As seen in Fig. 2B, the equivalent index increases exponentially with the radius and the equivalent grating period (which is constant in the real plane) also increases with the radius. This index profile does not necessarily support a guided defect mode.

In the ($U$, $V$) plane, the radial grating are transformed to a series of parallel grating normal to the $V$ axis but with an exponential index profile. In order for this structure to act as a Bragg reflector, the partial reflections from each interface must interfere constructively (see Fig. 3). In order for that to happen, the total phase that the wave accumulates while propagating through the layer should be $\pi/2$. This condition determines the layer width as follows:

$$\frac{\pi}{2} = \int k_{\perp} \cdot dU = \int \sqrt{k_0^2 n_{eq}^2 - \beta_V^2} \cdot dU \tag{13}$$

where the integration beginning and ending coordinates correspond the interfaces of the layer, $n_{eq} = \bar{n}\,\rho/R$ and $\bar{n} = n_j^2/\gamma_j$ according to equation (9). The propagation factor $\beta_V$ appearing in (13) is determined by the azimuthal wavenumber *m*:

$$\beta_V = m/R \tag{14}$$

Equation (13) was used to calculate the structures required for the high reflection Bragg mirrors surrounding the defect. Assuming the Bragg reflectors on both sides have identical reflection phase, then the defect width must be "$\lambda/2$" in the sense of eq. (13), i.e. the defect must satisfy:

$$l\pi = \int k_{\perp} \cdot dU = \int \sqrt{k_0^2 n_{eq}^2 - \beta_V^2} \cdot dU \qquad l = 1,2,3... \tag{15}$$

where the integer *l* indicates the number of the Bessel periods (or the radial modal number) of the field in the defect.

It follows that the width of the defect and Bragg layers depends on their coordinate *U* (or $\rho$) because the equivalent index $n_{eq}$ is a function of *U.*

Fig. 4 depicts the index (A) and the modal field (B) profiles of an annular defect mode resonator. The high index layers have effective refractive index ($\bar{n}$) of 2 while the low index layers and the defect have effective refractive index of 1. The internal and external Bragg reflectors have 5 and 10 periods respectively, the wavelength is 1.55µm and the azimuthal wavenumber is 7. The defect is located approximately at $\rho$=5.6µm and it is 0.85µm wide.

Fig. 5 shows the width of the high-index (stars) and low-index (circles) layers. At small radii the layers' width is wider because the equivalent index is lower there. The layers' width decreases for larger radii and approaches asymptotically the "conventional" quarter wavelength width - $\lambda/4n$. The two exceptionally wider low-index layers in Fig. 5 are the first low-index layer ($\rho$=0-2µm) and the defect which has a "$\lambda/2$" width.

**4. Modal solution properties**

Because of the design method ("$\lambda/4$" layers and "$\lambda/2$" defect) the resonator has a single radial mode which its peak is located almost exactly in the middle of the defect (see also Fig. 4). This is unlike the field profile of conventional ring resonators in which the field peak tends to shifts towards the exterior radius of the waveguide due to the increase on the equivalent $n_{wq}$ index in larger radii. Nevertheless, the asymmetry of the field profile (with respect to the intensity peak) which is due to the radial structure is noticeable.

Fig. 6 shows the dispersion curve of the annular defect resonator presented in Fig. 5. The vertical and horizontal axes indicate respectively the wavelength and the azimithal wavenumber *m*. The circles indicate the resonance wavelengths and the solid line represent a quadratic interpolation:

$\lambda = 1.6309 - 4.8 \cdot 10^{-3} \cdot m - 9.63 \cdot 10^{-5} \cdot m^2$. The resonator free spectral range (FSR)

around 1.55μm is approximately 20nm and it increases for shorter wavelengths. It is interesting to note that the quadratic term is the most dominant term in the determination of the resonator FSR.

Figure 7 depicts the transverse profile of the modal fields corresponding to changing the azimuthal wavenumber from $m$=6 to 10. It is evident that the transverse profile is almost identical although the resonance wavelength changes over more than 100nm. The reason for that is that transverse profile is primarily determined by the Bragg layers widths (or spatial frequency) which is wavelength independent. This feature is an important advantage compared to conventional ring resonators because coupling between resonators of this type and Bragg waveguides, which is determined primarily by the modal profiles overlap, can be expected to be almost wavelength independent.

Figs. 8, 9 show the dispersion and the transverse profiles of $m$=6 to 12 for a Bragg defect resonators utilizing lower refractive index contrast. For this structure, $\overline{n}_{core} = \overline{n}_2 = 3.0$ and $\overline{n}_1 = 3.5$, the internal and external Bragg layers have both 40 periods and it was designed for $m$=10 at 1.55μm. The defect is located at $\rho$=10.85μm and its width is approximately 0.27μm. Because of the lower contrast, more Bragg layers are needed in order to realize good mode confinement and, as a result, the resonator is larger and the FSR is smaller (about 96GHz at 1.55μm).

As shown in Fig. 8, the dispersion curve of this resonator is also quadratic and given by $\lambda = 1.5541 - 1.2 \cdot 10^{-5} \cdot m - 3.98 \cdot 10^{-5} \cdot m^2$. Similar to the high-contrast case, the modal transverse profile exhibits small wavelength dependence, which can be primarily seen in the small radii regime.

## 5. Higher order Bragg reflectors

Although the chirped quarter wavelength Bragg layers form an optimal reflector, their implementation could prove to be difficult. Because the layers' spatial period changes, some of the conventional photolithography methods which are employed for uniform (not chirped) Bragg gratings [20] cannot be used. A possible approach to overcome this problem is to position the interfaces in non-sequential zeros/extrema, i.e. allow the Bessel function in each layer to complete a full period before changing the index. From the Bragg reflection point of view, such approach is equivalent to utilizing $(2l+1)\lambda/4$ layers or employing higher reflection order of the Bragg stack. Practically, the layer width can be evaluated in a similar fashion to the quarter wavelength structure but the layers have to satisfy the following condition:

$$(2l+1)\frac{\pi}{2} = \int k_{\perp} \cdot dU = \int \sqrt{k_0^2 n_{eq}^2 - \beta_V^2} \cdot dU \qquad l = 1,2,3... \tag{16}$$

The resulting structure would have wider layers and would, therefore, be larger and exhibit smaller FSR. Fig. 10 compares between the field transverse profile of Fig. 4 (A) and the transverse profile of a resonator designed for similar mode parameters (5 internal periods, 10 external periods *m*=7 for $\lambda$=1.55μm) utilizing wider layers (B). The radius at which the field amplitude peaks is more than twice larger than the original (11.35μm Vs. 5.6μm) and the radial decay of the field rate is smaller. Fig. 11 depicts the dispersion curve of the higher order Bragg based resonator. As expected, the FSR of the resonator is significantly smaller than that of the original one (approximately 3nm at 1.55μm). The main reason for this decrease in the FSR is the increase in the defect radius. However, since the lower index layers are inherently wider (especially in the lower radii regime), they could be realized utilizing quarter wavelength layers while the high-index would be realized as a higher order Bragg layer. Moreover, depending on the radius and the index, the Bragg order of each layer could be determined separately to achieve the largest FSR.

Fig. 12 depicts the dispersion curve of a ring resonator designed for the same parameters as those shown in Figs. 4 and 9 where the low-index layers are first order Bragg layers and the high-index layers are second order Bragg layers. The implementation of the composite structure increased the FSR from 3nm to ~8nm without requiring smaller features. As for the other Bragg defect resonators shown here, the modal transverse profile of this resonator is almost wavelength independent.

**6. Discussion and Summary**

We have analyzed annular defect mode ring resonator based of radial Bragg reflectors. We also presented a simple and intuitive method to design the reflection gratings of such resonators. We saw that extremely small resonators (few microns in diameter) exhibiting large FSR can be realized utilizing relatively low index materials.

Several configurations to realize the Bragg reflectors were suggested and analyzed. The straightforward configuration (each layer serves as an equivalent quarter wavelength plate) offers the smallest resonator exhibiting the largest FSR. On the other hand, manufacturing of such device may require the realization of small and accurate features, especially if the optimal gratings structure is required. Employment of higher-order Bragg gratings relaxes the tolerances on the gratings width but deteriorates the FSR.

The composite configuration, i.e. tailoring each layer Bragg order and width according to its refractive index and radius, seems to be the best compromise between large FSR and realizable features. Quarter wavelength layers can be easily realized if the material refractive index is low or if the layer is positioned in a small radius where the equivalent index, $n_{eq}$, is low. Employing the thinnest possible Bragg layers is important especially for the internal Bragg reflector because this would determine the defect radius and hence, the FSR. The external Bragg reflector could be realized using higher-order Bragg layers without a major influence on the resonator performances. The more tolerant higher Bragg order resonators design enables a relatively simple realization by conventional photolithography techniques in a variety of material systems such as GaAs, InP and polymers.

**6. Acknowledgments**

The authors thank Shayan Mookherjea and George T. Paloczi for useful discussions. This work was supported by the U.S. Office of Naval Research (Y. S. Park) and the U.S. Air Force Office of Scientific Research (H. Schlossberg).

Figure Caption

Figure 1 – An illustration of the annular defect mode resonator structure.

Figure 2 – The radial refractive index profile (A) and the equivalent index profile (B) of an annular defect surrounded by Bragg reflectors. The maximal and minimal refractive index is 1.5 and 1 respectively and the grating period is ~1μm.

Figure 3 – An illustration of the design rule used realize a highly efficient radial Bragg reflector.

Figure 4 – Radial index profile (A) and electrical field distribution (B) of an annular defect mode resonator.

Figure 5 – The high-index (stars) and low-index (circles) layers widths of the resonator shown in figure 4.

Figure 6 – Resonance wavelengths (circles) and a quadratic fit (solid line) of the resonator shown in figure 4.

Figure 7 – Modal field profiles for $m$=6 (dotted line), 7 (solid line) and 10 (dash-dotted line) of the resonator shown in figure 4.

Figure 8 – Resonance wavelengths (circles) and a quadratic fit (solid line) of a resonator based on lower contrast Bragg reflectors.

Figure 9 – The modal field profile of the resonator based on lower contrast Bragg reflectors.

Figure 10 – A comparison of the modal field profile shown in figure 4 (A) and the modal field of a resonator based on second-order Bragg reflectors with similar parameters (B).

Figure 11 – Resonance wavelengths (circles) and a quadratic fit (solid line) of the second-order Bragg reflectors based resonator.

Figure 12 – Resonance wavelengths (circles) and a quadratic fit (solid line) of a resonator based on composite Bragg reflectors with similar parameter to the resonator of Fig. 4. The low-index layers are quarter wavelength in width and the high-index layers are three-quarter wavelength in width.

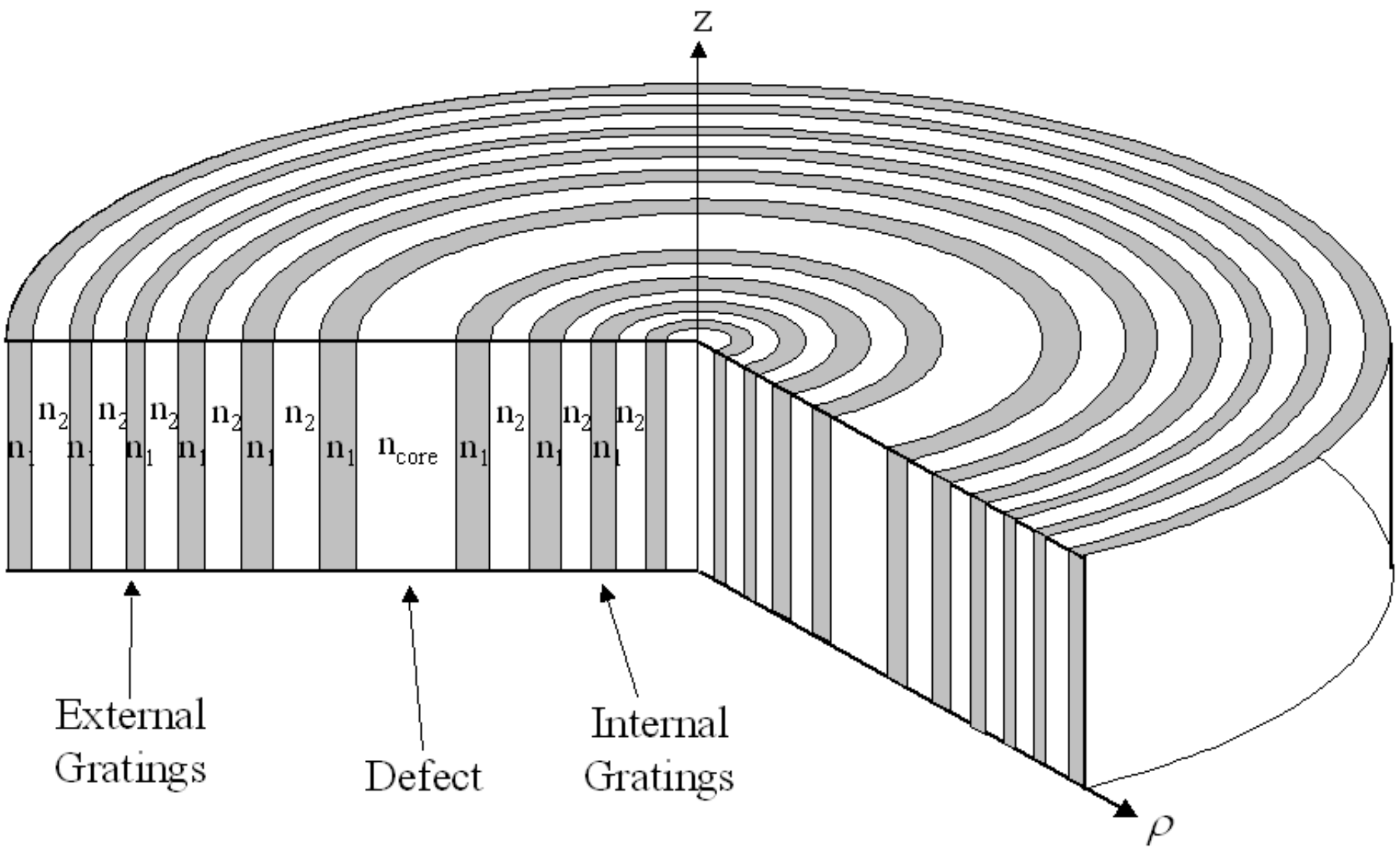
z
$n_2$ $n_2$ $n_2$ $n_2$ $n_2$ $n_2$ $n_2$ $n_2$
$n_1$ $n_1$ $n_1$ $n_1$ $n_1$ $n_1$ $n_{core}$ $n_1$ $n_1$ $n_1$
External Gratings
Defect
Internal Gratings
$\rho$

Figure 1

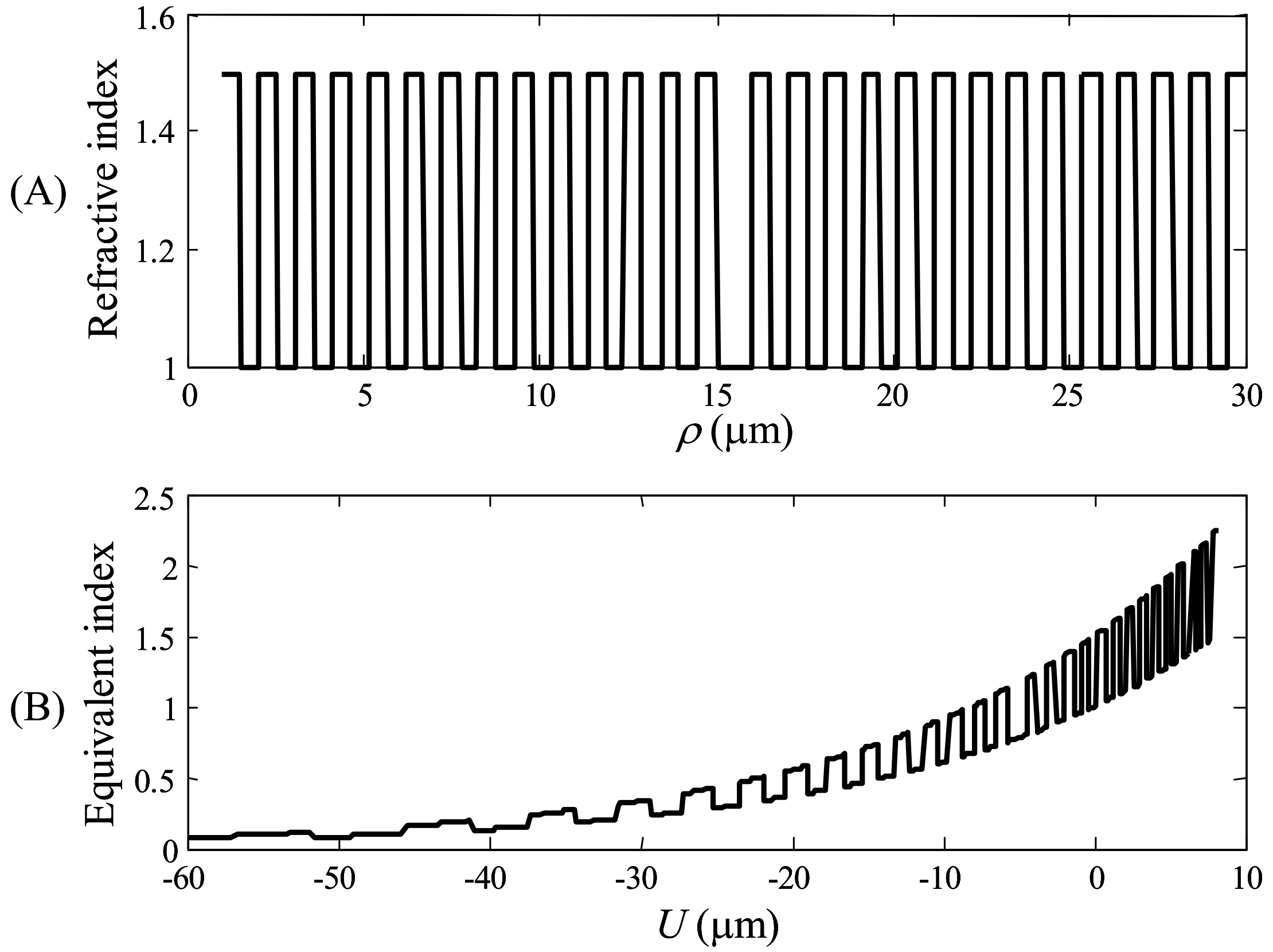

Figure 2

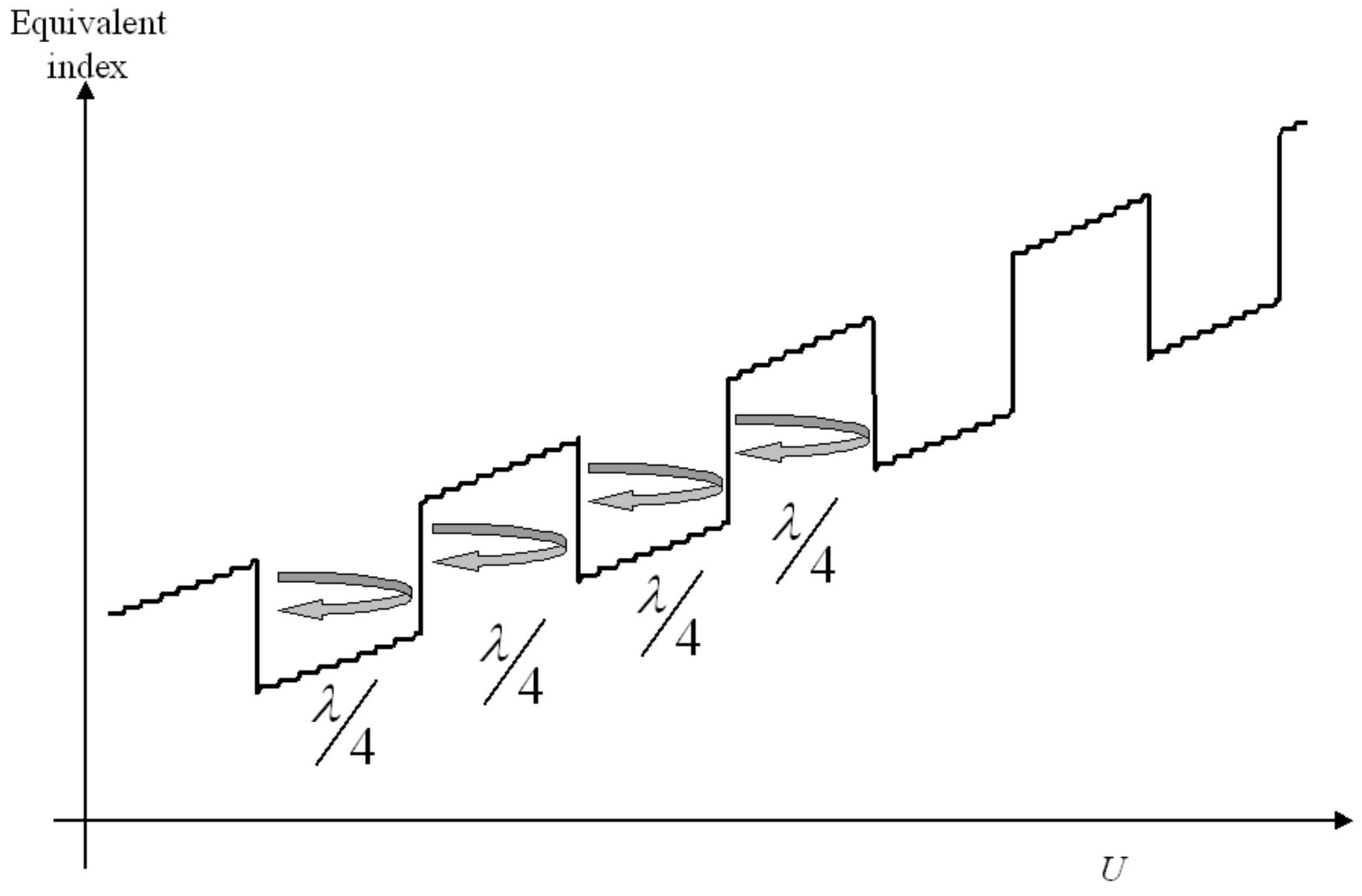

Figure 3

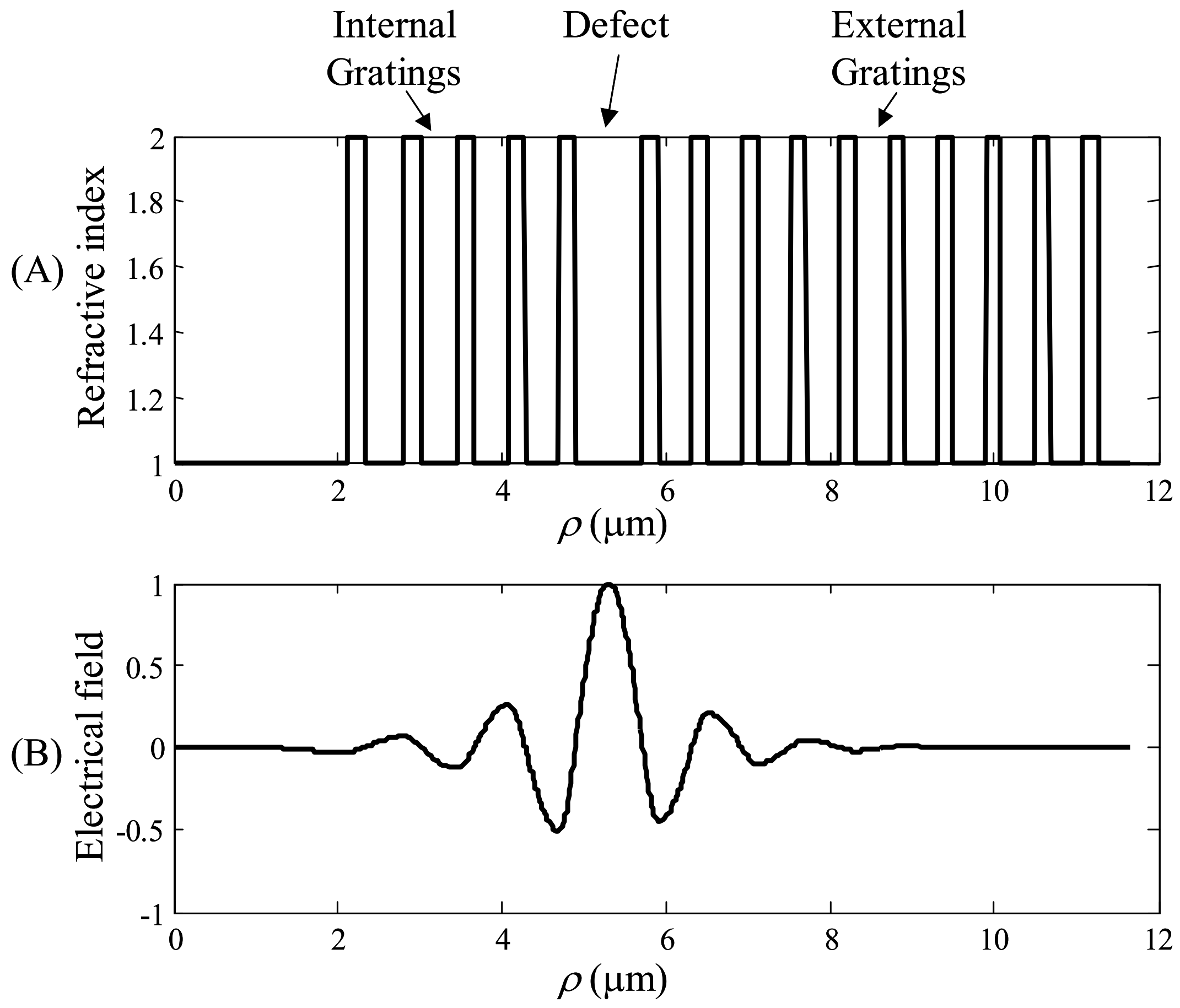

Figure 4

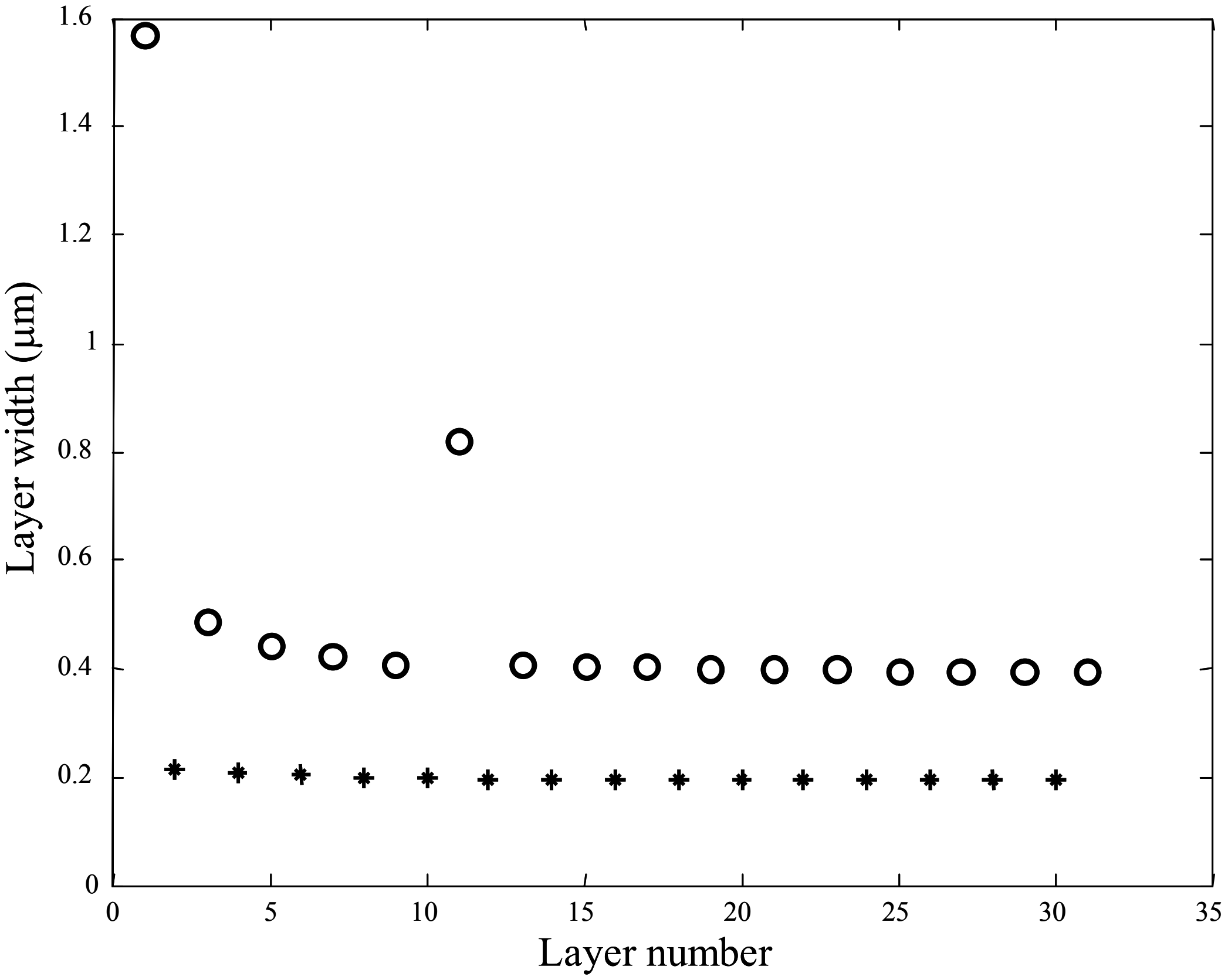

Figure 5

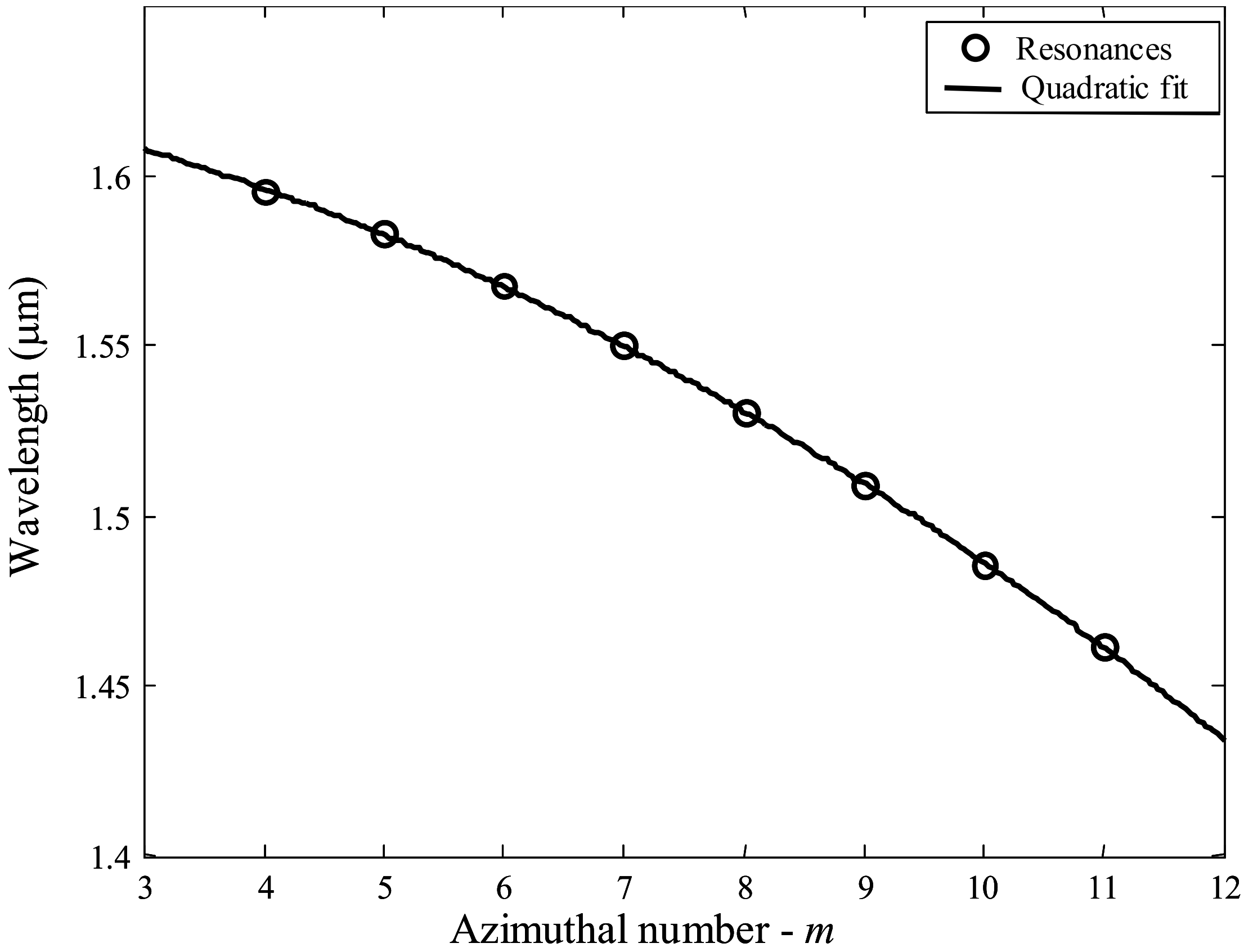

Figure 6

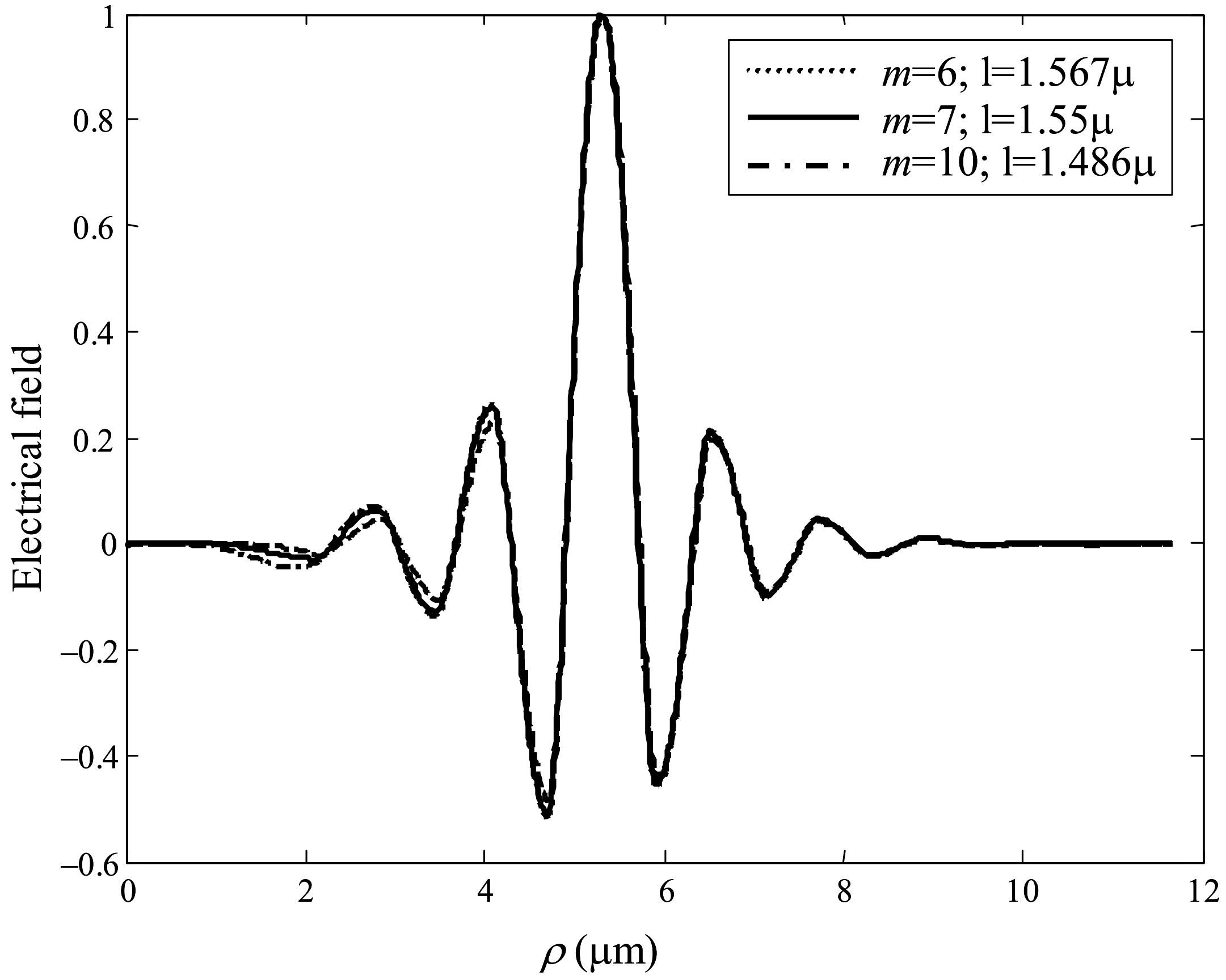

Figure 7

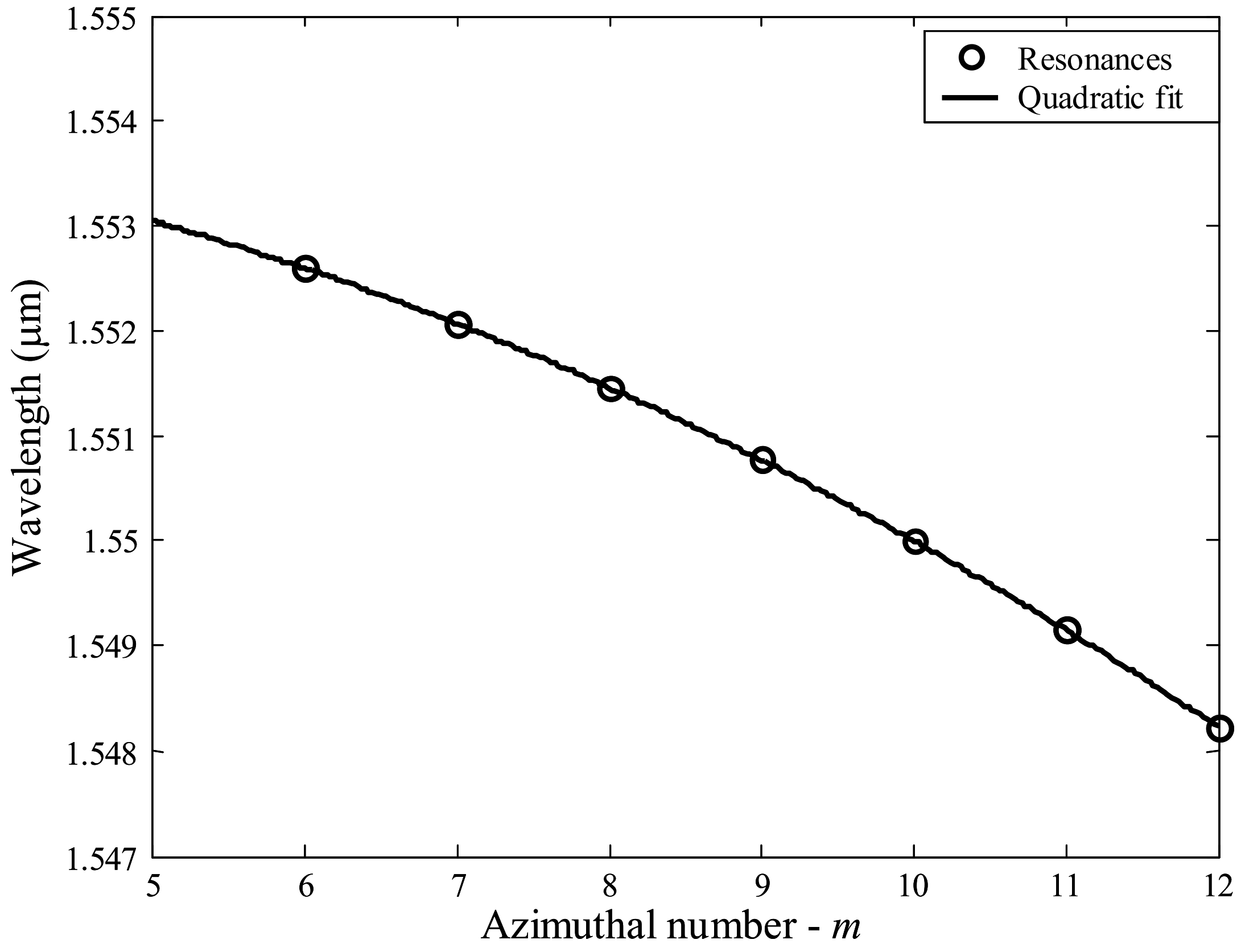

Figure 8

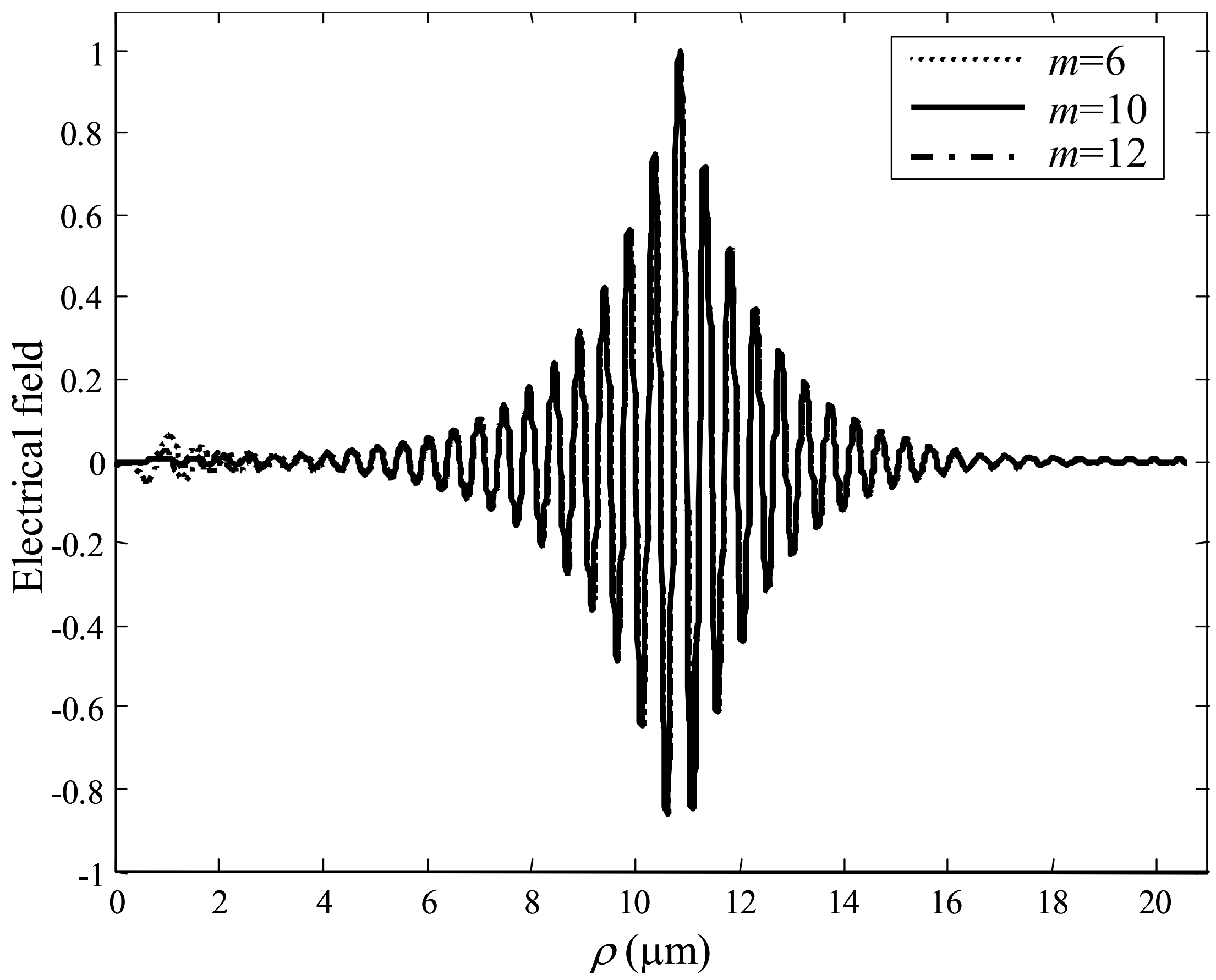

Figure 9

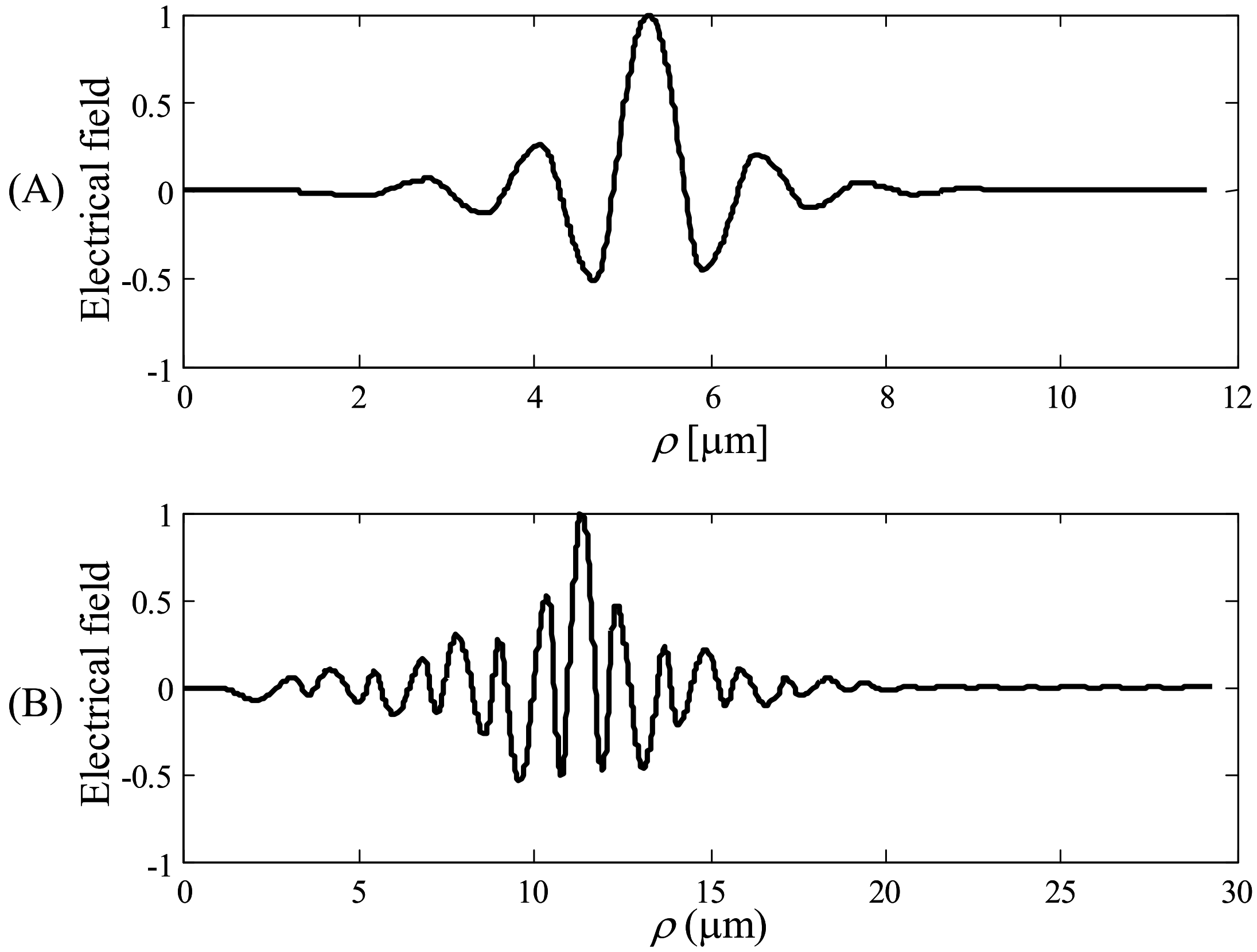

Figure 10

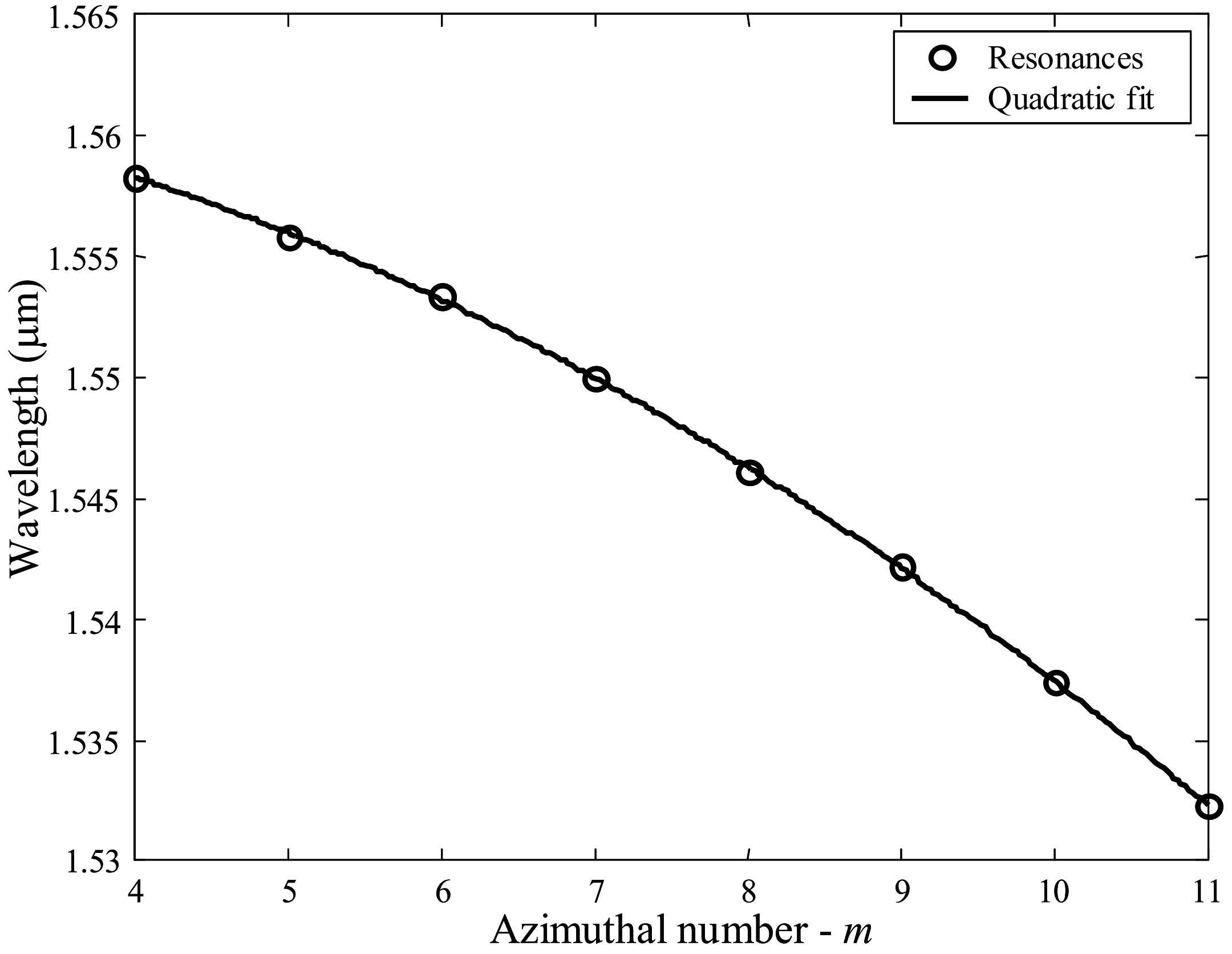

Figure 11

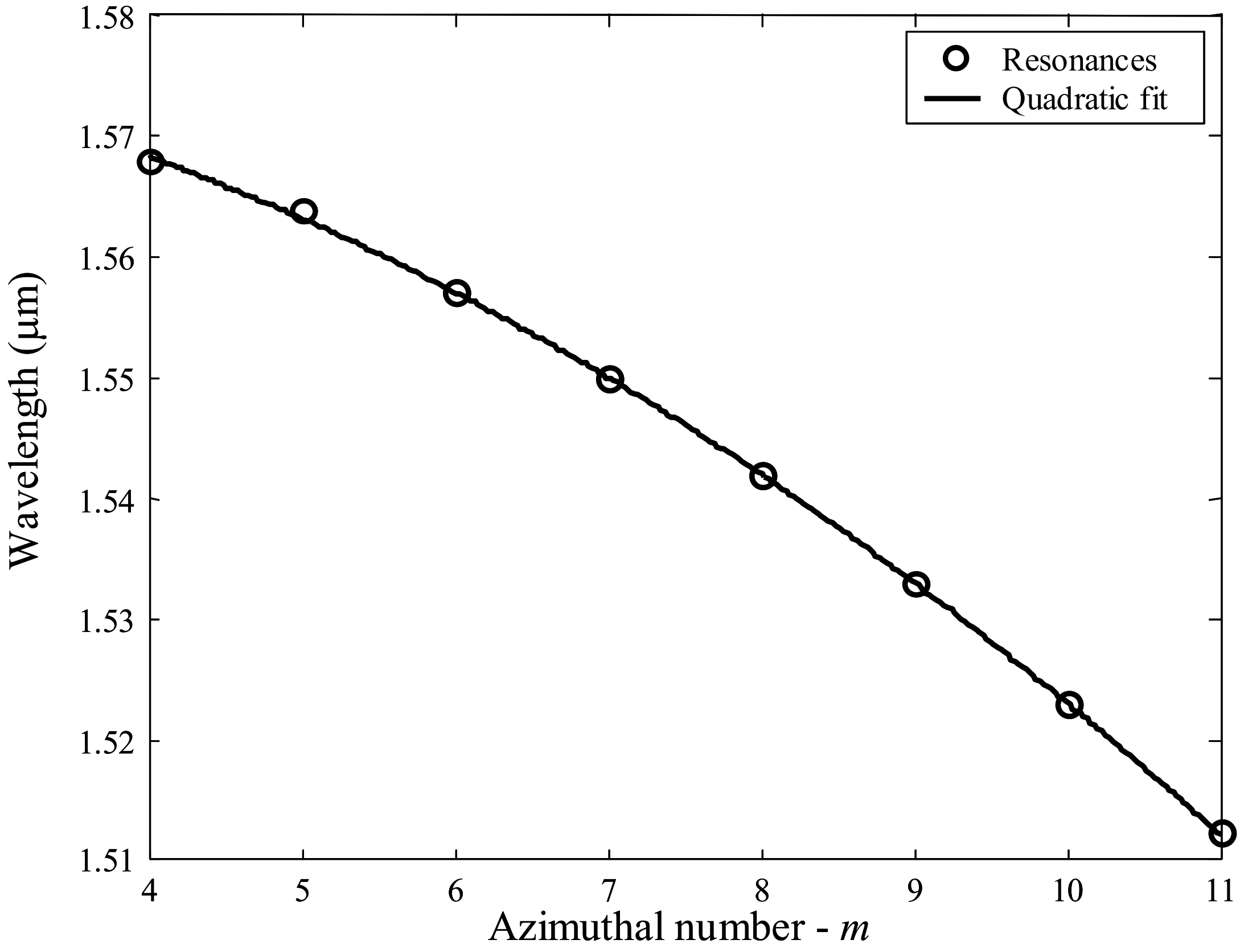

Figure 12